\documentclass[onecolumn]{aastex631}

\usepackage{graphicx}
\usepackage{amssymb}
\usepackage{amsmath}
\usepackage{epstopdf}
\usepackage{gensymb}
\usepackage{natbib}
\usepackage{appendix}
\usepackage{tabu}
\usepackage{color}

\defcitealias{Nina1}{N20}
\defcitealias{Isik2018}{I18}
\defcitealias{Shapiro2014}{S14}

\accepted{\today}

\shortauthors{N\`emec et al.}

\graphicspath{{./}{figures/}}

\begin{document}

\title{Faculae cancel out on the surfaces of active Suns}
\shorttitle{}

\correspondingauthor{N.-E.~N\`emec}
\email{nina-elisabeth.nemec@uni-goettingen.de}

\author[0000-0001-6090-1247]{N.-E.~N\`emec}
\affiliation{Institut für Astrophysik und Geophysik, Georg-August-Universität Göttingen, Friedrich-Hund-Platz 1, D-37077 Göttingen,Germany}
\affiliation{Max-Planck-Institut f\"ur Sonnensystemforschung, Justus-von-Liebig-Weg 3, 37077 G\"ottingen, Germany}

\author[0000-0002-8842-5403]{A.~I.~Shapiro}
\affiliation{Max-Planck-Institut f\"ur Sonnensystemforschung, Justus-von-Liebig-Weg 3, 37077 G\"ottingen, Germany}

\author[0000-0001-6163-0653]{E.~I\c{s}{\i}k}
\affiliation{Department of Computer Science, Turkish-German University,
\c{S}ahinkaya Cd. 94, Beykoz, 34820 Istanbul, Turkey}

\author[0000-0002-3243-1230]{K. Sowmya}
\affiliation{Max-Planck-Institut f\"ur Sonnensystemforschung, Justus-von-Liebig-Weg 3, 37077 G\"ottingen, Germany}

\author[0000-0002-3418-8449]{S.~K.~Solanki}
\affiliation{Max-Planck-Institut f\"ur Sonnensystemforschung, Justus-von-Liebig-Weg 3, 37077 G\"ottingen, Germany}
\affiliation{School of Space Research, Kyung Hee University, Yongin, Gyeonggi 446--701, Korea}

\author[0000-0002-1377-3067]{N~.A.~Krivova}
\affiliation{Max-Planck-Institut f\"ur Sonnensystemforschung, Justus-von-Liebig-Weg 3, 37077 G\"ottingen, Germany}

\author[0000-0001-9474-8447]{R.~H.~Cameron}
\affiliation{Max-Planck-Institut f\"ur Sonnensystemforschung, Justus-von-Liebig-Weg 3, 37077 G\"ottingen, Germany}

\author[0000-0001-7696-8665]{L.~Gizon}
\affiliation{Institut für Astrophysik und Geophysik, Georg-August-Universität Göttingen, Friedrich-Hund-Platz 1, D-37077 Göttingen,Germany}
\affiliation{Max-Planck-Institut f\"ur Sonnensystemforschung, Justus-von-Liebig-Weg 3, 37077 G\"ottingen, Germany}
\affiliation{Center for Space Science, NYUAD Institute, New York University Abu Dhabi, Abu Dhabi, UAE}

\begin{abstract}
Surfaces of the Sun and other cool stars are filled with magnetic fields, which are either seen as dark compact spots or more diffuse bright structures like faculae. Both hamper detection and characterisation of exoplanets, affecting stellar brightness and spectra, as well as transmission spectra. However, the expected facular and spot signals in stellar data are quite different, for instance they have distinct temporal and spectral profiles. Consequently, 
corrections of stellar data for magnetic activity can greatly benefit from the insight on whether the stellar signal is dominated by spots or faculae.  Here, we utilise a surface flux transport model (SFTM) to show that more effective cancellation of diffuse magnetic flux associated with faculae leads to spot area coverages  increasing faster with stellar magnetic activity than that by faculae. Our calculations explain the observed dependence between solar spot and facular area coverages and allow its extension to stars more active than the Sun. This extension enables anticipating the properties of stellar signal and its more reliable mitigation, leading to a more accurate characterisation of exoplanets and their atmospheres.
\end{abstract}

\keywords{Stellar rotation (1629) -- Stellar activity (1580))}

\section{Introduction}
The magnetic fields emerging on solar and stellar surfaces lead to the formation of surface magnetic features, such as dark spots and bright faculae \citep[see, e.g.,][for a review of the solar case]{Sami_rev2006}. An exciting manifestation of these magnetic features is the temporal variation of brightness of the Sun and solar-like stars \citep{Solanki2013,Ermolli2013,Kopp_SP}. 

The most precise records of solar brightness are obtained by space-borne radiometers which measure the Total Solar Irradiance (TSI), i.e. the wavelength-integrated solar radiative flux normalised to a distance of 1 AU \citep[see ][for review]{Kopp2014}. These measurements of the TSI revealed that it varies along the course of the 11-year sunspot activity cycle. 
Namely, the TSI is about 0.1\% higher at the solar maximum than at the solar minimum. This implies that the decrease of the TSI due to spots is overcompensated by the increase caused by faculae, i.e. the TSI variability on the activity cycle timescale is facula-dominated \citep[see e.g. the detailed discussion and references in][]{Shapiro2016}. 

Concurrently to solar studies, a significant effort was invested in studying magnetic activity cycles and brightness variations of other stars. Long-term monitoring programs of stellar variability began in the 1960's, as Olin Wilson established the Mount Wilson observations of the Ca II H and K lines. The emission in  Ca II H\&K lines formed in stellar chromosphere is a good proxy of magnetic heating, so that the Mount Wilson program allowed studying variations of stellar magnetic activity. The next big step in observations of stellar variability became the establishment of the Lowell Observatory monitoring program \citep{Lockwood1992}. This program conducted observations in
the visual spectral domain in the \textit{Strömgren b} and \textit{y} filters, which are centred at wavelengths of 467 and 547 nm, respectively. Simultaneous measurements of stellar photometric brightness and chromospheric activity revealed a distinction between active young and less active old stars \citep[e.g.][]{Lockwood1997,Radick1998,Radick2018,Lockwood2007}. It turned out that the Ca II line emission and the \textit{Strömgren b} and \textit{y} brightness are positively correlated (i.e. the increase in the Ca II emission is accompanied by an increase in the photometric brightness, like for the Sun) for less active stars and negatively correlated in more active stars.

\cite{Shapiro2014} showed that this transition from faculae- to spot-dominated stellar brightness variability can be explained by a simple extrapolation of solar observation that  faculae-to-spot area ratio decreases with increasing activity level \citep{Foukal1993,Chapman1997,Foukal1998}. An intriguing application of this result is 
that it provides evidence that the relation between facular and spot disk coverages observed on the Sun is valid, at least qualitatively, also for other solar-like stars, and, in particular, for stars more active than the Sun. This raises the question about the physical mechanism behind this relation.


In this study we employ the surface flux transport model \citep[in the form of][]{Isik2018} to model the emergence and evolution of flux on the surfaces of stars. Then we obtain the area coverage of faculae and spots following the approach presented in \cite{Nina1} (hereafter \citetalias{Nina1}). In Sects.~ 2.1~and~2.2. we demonstrate that our model allows reproducing both solar and stellar data. In Sect.~3 we then use our model to explain what leads to the decrease of the faculae-to-spot ratio with magnetic activity.

\section{Facular and spot disc coverages: model vs. observations}

\subsection{The Sun}
We first considered a single cycle on a star with the activity level corresponding to solar cycle 22 (1986--1996). For this we applied a record of  active region emergences in the form of bipolar magnetic regions (BMRs) calculated from a semi-empirical model of the solar cycle \citep{Jiang2011_1}  and ran the SFTM to obtain daily surface distributions of magnetic field.
 A detailed description of the SFTM setup is given in Appendix A. 

\begin{figure}[h!]
\centering
\includegraphics[width=0.45\textwidth]{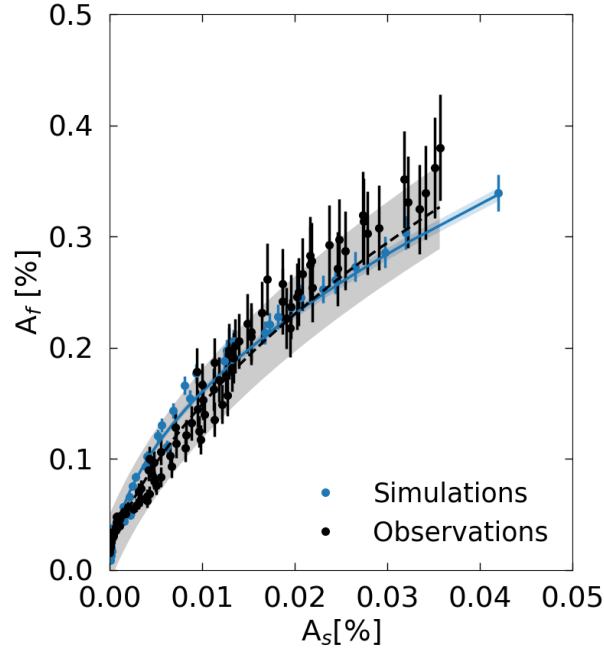}
\caption{The dependence of the facular area coverage on that of spots as returned by our forward model (blue) and retrieved from solar observations (black). Dots represent 58-day binned values (see text), the vertical lines represent the standard deviations in the corresponding bins. The solid curves represent least-square fits with the function $A_f = a\cdot \sqrt{A_s+c}+d$, and 
shaded areas are the 1-$\sigma$ uncertainties of the fits.}
\label{fig:fig_1}
\end{figure}

We then followed the approach developed in \citetalias{Nina1} to transform these distributions to maps of visible disc distribution of faculae and spots. In Fig.~\ref{fig:fig_1} we plot the forward-modelled relation
between the disc area coverage of faculae, $A_f$, and that of spots, $A_s$, for a star with activity level corresponding to solar cycle 22 and compare it to solar observations collected by \cite{Shapiro2014}. The daily values of $A_f$ and $A_s$ were both sorted according to ascending $A_s$, before binning them in 58-day intervals to remove the effect of solar rotation. This interval is the same as in Shapiro et al. (2014) and ensures that effects that are purely due to transits of features (i.e. due to the solar rotation) are removed. Longer intervals than several solar rotations would on the other hand remove the effects of the evolution of the magnetic features, which are of interest in the present work. We have checked that the result it not affected by reasonable choices of binning intervals (i.e. over 1 or 3 solar rotations.) Fig.~\ref{fig:fig_1} demonstrates that our calculations reproduce the observed trend of facular area increasing less rapidly with magnetic activity than the spot area.

\begin{figure}[h!]
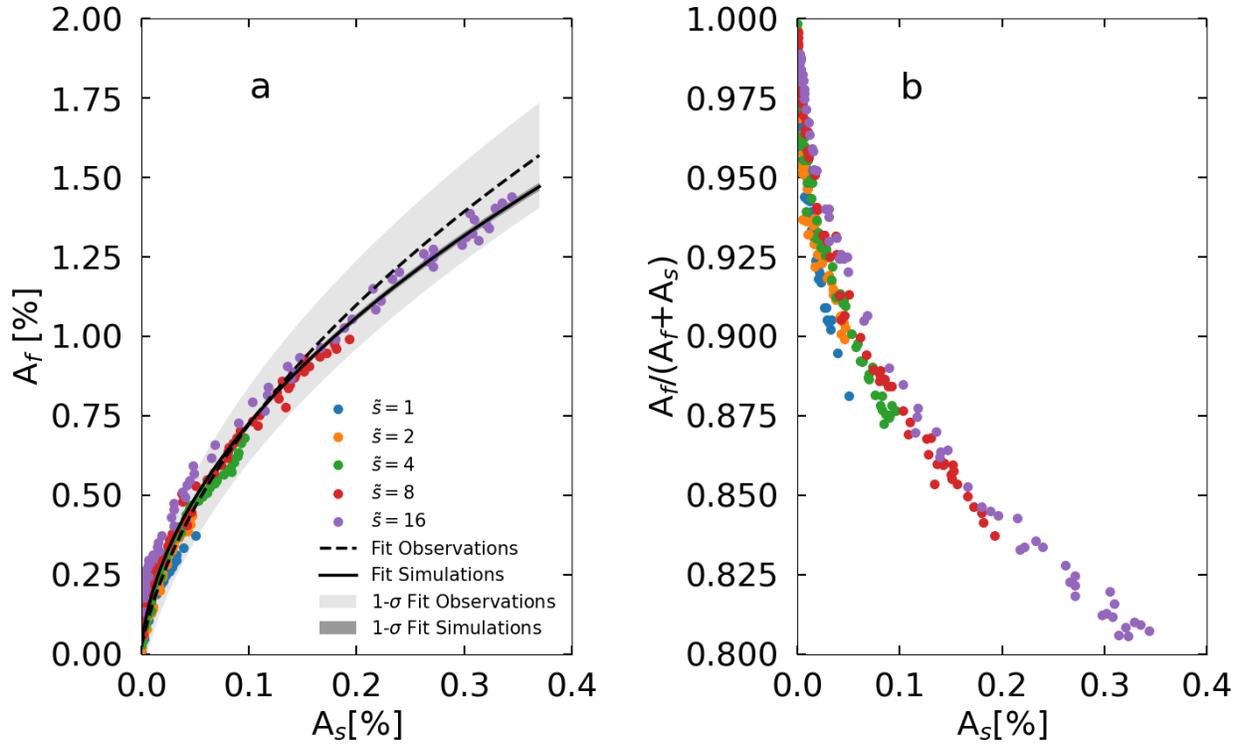

\centering
\includegraphics[width=0.45\textwidth]{Figure_2a.pdf}\quad
\includegraphics[width=0.45\textwidth]{Figure_2b.pdf}
\caption{Similar to Fig. \ref{fig:fig_1} but the forward-modelled dependence is plotted for stars with different levels of magnetic activity.  Panel b gives the fraction of faculae in total area of active regions as a function of the spot area.}
\label{fig:fig_2}
\end{figure}

\subsection{Active stars }
We now extend the model to stars more active than the Sun before unveiling the physical mechanisms responsible for this curios behaviour of facular and spot coverages. We model active stars by filling their surfaces with gradually increasing amounts of magnetic flux. This is achieved by scaling  the number of emergences by a given factor, $\tilde{s}$ \citep[see][and Appendix B]{Isik2018}, and running the SFTM model again. A star with, for example, $\tilde{s}=$2 experiences twice as many emergences than the solar case  $\tilde{s}=$1.

The resulting facular and spot areas are shown Fig.~\ref{fig:fig_2}a, where we plot together calculations representing the Sun during cycle 22 ($\tilde{s}=1 $), more active stars ($\tilde{s}=2,\,4,\,8,\,16$, see Appendix B), and the fits to the  solar area coverages from Fig.~\ref{fig:fig_1}, which were  extrapolated towards higher spot areas.
Fig.~\ref{fig:fig_2}b additionally gives the ratio of facular area coverage to total area coverage by magnetic features (i.e. the sum of faculae and spots) to detail the decrease in the facular component of magnetic activity. All in all, Figs. \ref{fig:fig_2}a,b  indicate that the decrease of the facular-to-spot ratio with activity continues also well beyond the observed level of solar activity.

While comparing our results to solar data was relatively straightforward, only some indirect measurements of spot coverages of solar-like stars are available and there is basically no reliable technique for measuring facular coverages. Thus, we cannot directly test the dependencies plotted in Fig.~\ref{fig:fig_2} against stellar observations. Instead we opted to investigate whether the modelled trends in disc coverages by magnetic features are consistent with the observed variability of stellar brightness along the course of stellar activity cycles. 
In particular, synoptic programs at the Lowell and Fairborn observatories \citep{Lockwood1992, Lockwood2007, Radick1998, Radick2018} revealed that an increase of stellar activity (quantified via the S-index, which is a measure of stellar emission in the Ca II H\&K lines and is routinely used as a proxy of stellar magnetic activity) in time is accompanied by an increase in the stellar photometric brightness in Strömgren (b+y)/2 for stars with similar activity levels as that of the Sun. However, stars with higher activity levels show an anti-correlation between magnetic activity and photometric brightness (i.e. the star gets photometrically darker with increasing Ca II H\&K  emission).

\begin{figure}[h!]
    \centering
    \includegraphics[width=0.5\columnwidth]{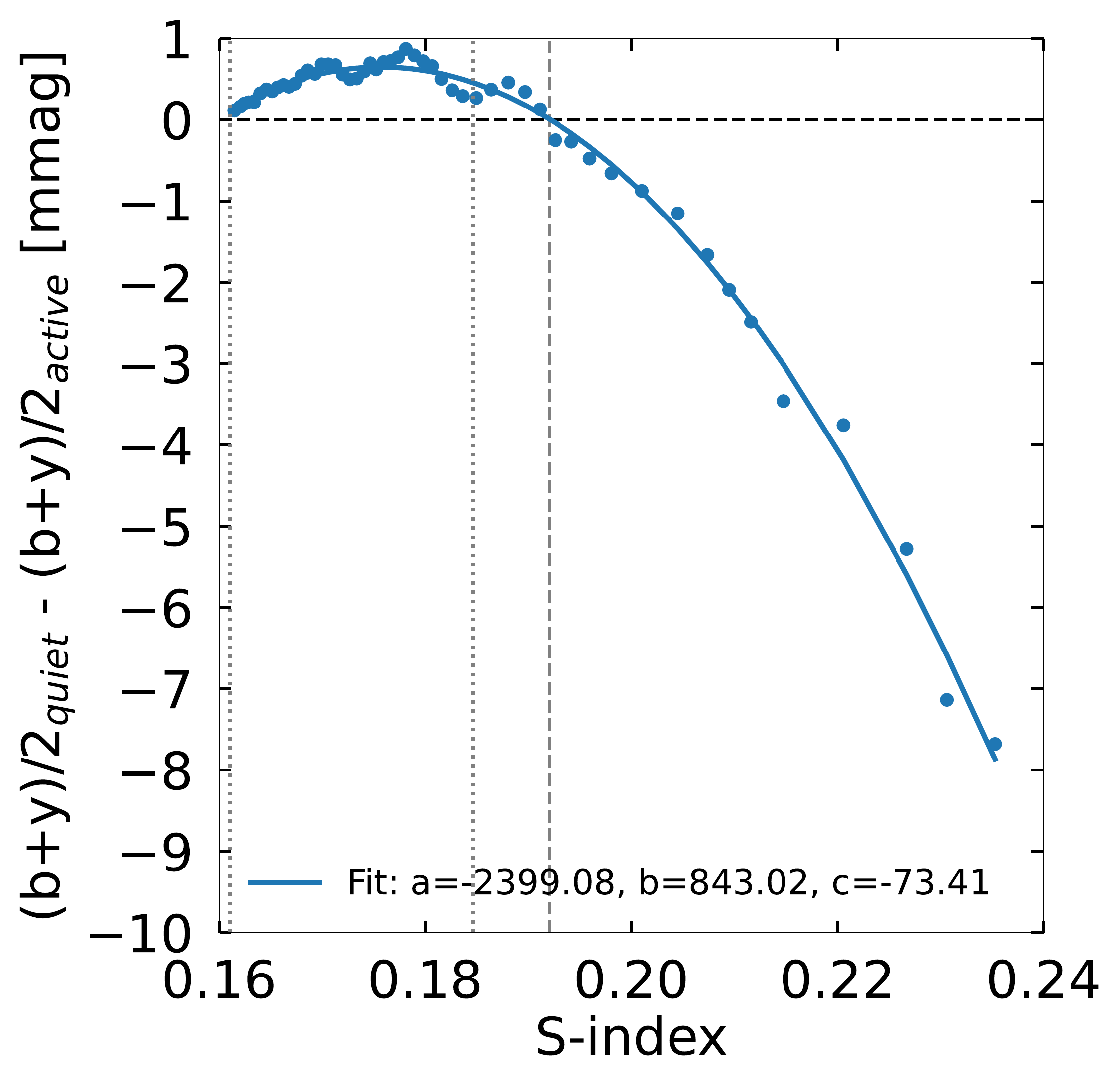}
    \caption{Magnetic alteration of stellar brightness in Strömgren (b+y)/2 as a function of  S-index.  Plotted are 365-day binned values (blue dots, see text) and standard deviations of brightness within each of the bins (vertical blue bars, which are mostly smaller than the dot size). The solid blue line gives the fit of the second-degree polynomial in the form of $aS^2+bS+c$, with the coefficients of the fit shown in the legend. The horizontal black dashed line indicates the zero level and the vertical grey dashed line points to the S-index value, at which the transition from faculae-to-spot magnetic alteration occurs. Vertical dotted grey lines indicated the minimum and maximum value of the calculations representing the Sun.
    } 
    \label{fig:poly-fit}
\end{figure}

Previous studies indicated that such a transition can be reproduced by extrapolating the observed dependencies of solar facular and spot area coverage on the S-index \citep{Shapiro2014}. Instead of relying on these extrapolations, which are purely empirical and assume that any trend seen on the Sun continues unchanged to more active stars, we combined our calculations of spot and facular coverages with models of stellar photometric brightness from \citetalias{Nina1} and Ca II H\&K  emission developed by \cite{Sowmyaetal2021}. Namely, we used the model introduced in \citetalias{Nina1} to calculate the activity-induced alteration in Strömgren (b+y)/2, $(b+y)/2_{\rm quiet}-(b+y)/2_{\rm active}$, where the first term represents the brightness of a star free from any magnetic features and the second term quantifies the brightness in the presence of magnetic activity (so that a positive alteration of Strömgren (b+y)/2 corresponds to an overall brightening with increasing magnetic activity, while a negative alteration corresponds to a dimming). These calculations have been performed for the daily snapshots of the disc distributions of magnetic features for all activity levels ($\tilde{s}$) used in this study (see Fig~{\ref{fig:fig_1}}b,c).
Then we used the \cite{Sowmyaetal2021} model for the Ca II H\&K  emission to calculate the corresponding S-index values. As a result, we obtained a contemporaneous time series of Strömgren (b+y)/2 and the S-index, by combining all five modelled activity levels into a 55-year time series. We then sorted both time series according to ascending S-index values, before calculating values averaged over bins containing 365 sorted data points  shown in Fig.~\ref{fig:poly-fit}. One can see that the alteration in Strömgren (b+y)/2 varies with the S-index, following a second-degree polynomial, and changing its sign at an S-index value of about 0.195, which is only slightly above the maximum of 0.188 observed on the Sun \citep{Shapiro2014}.

\begin{figure}[h!]
    \centering
    \includegraphics[width=0.65\columnwidth]{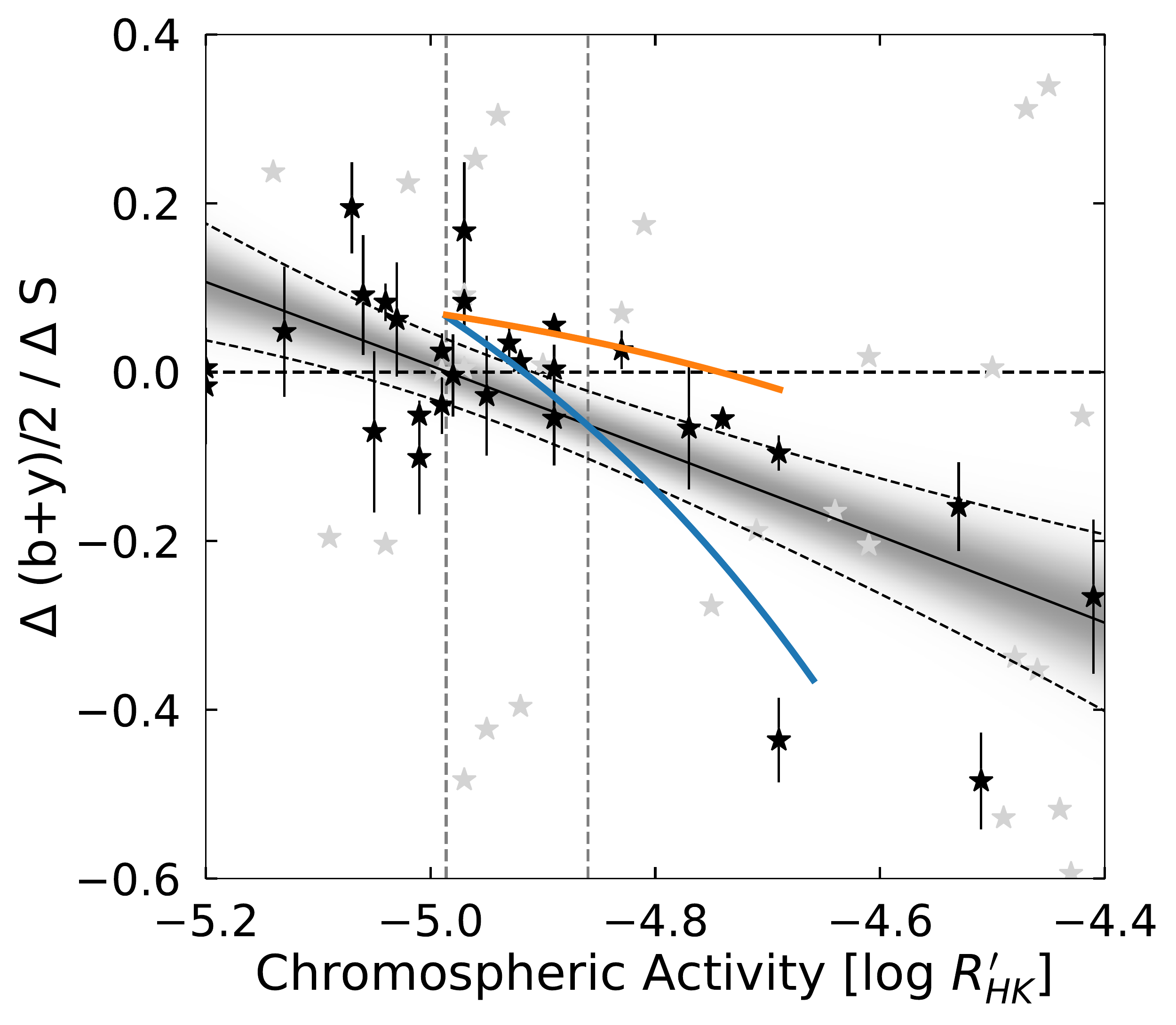}
    \caption{$\Delta[(b+y)/2]/\Delta S$ plotted against the mean chromospheric activity expressed through $ \log R'_{HK}$. The blue and orange solid lines represent the 
    modelled dependencies for $i=90\degree$ and $i=0\degree$, respectively, while asterisks show the Lowell and Fairborn measurements\citet{Radick2018}. Black asterisks  indicate stars with effective temperatures $\pm$200 K around the solar value (5777 K) and with uncertainty in $\Delta[(b+y)/2]/\Delta S$ (as indicated by the black bars)  below 0.1. Grey asterisks depict the rest of the Lowell and Fairborn sample.  The dashed horizontal line indicates the zero level, where the transition between faculae-to-spot dominated occurs, and the two vertical grey dashed lines indicate the minimum and maximum value of solar activity during cycle 22. The grey-shaded band shows the result of the Bayesian linear regression to the black asterisks using Gaussian priors for the parameters of a quadratic polynomial. The solid curve shows the maximum of the posteriori estimate and the shaded region represents the probability density of the posterior distribution. The dashed curves denote the $2\sigma$ posterior boundaries.}
    \label{fig:Radick_comp}
\end{figure}

Similar to the measurements from the  Lowell and Fairborn programmes, we calculated the slope between the changes of stellar photometric brightness and the S-index over the course of the magnetic activity cycle, $\Delta[(b+y)/2]/\Delta  S$. In this metric, positive values correspond to facula-dominated photometric variability as observed on the Sun and inactive stars. Negative values conversely correspond to spot-dominated photometric variability as observed on active stars. In Appendix C we show how to transform the dependence plotted in Fig.~\ref{fig:poly-fit} into the dependence of $\Delta[(b+y)/2]/\Delta  S$ on the stellar activity level.  Fig.~\ref{fig:Radick_comp} presents the comparison of our model output. As stars are observed from various, mostly unknown inclination angles  $i$ (i.e. angles between the direction to the observer and the stellar rotation axis), we show the calculations performed for two extreme cases: $i=90\degree$ (i.e. stars observed equator-on) and $i=0\degree$ (i.e. stars observed pole-on) in comparison to the Lowell and Fairborn data \citep{Radick2018}.We note that the inclination affects the variability by changing the disk distribution of magnetic features. For example, on stars like the Sun the magnetic features are mainly concentrated in the near-equatorial bands so that the change of the inclination from $i=90\degree$ to $i=0\degree$ will shift the magnetic features towards the (visible) limb, affecting both photometric variability \citep[see][]{Nina1} and the S-index  \citep[see][]{Sowmyaetal2021}.
In Fig.~\ref{fig:Radick_comp} we show stars with values of effective temperatures $\pm$200 K around the solar effective temperature. Such a choice is motivated by the strong dependence of facular and spot brightness contrast (and, thus, the regime of the variability) on stellar effective temperature \cite[see, e.g.,][]{Witzke2018, Witzke2020, Panjaetal2020}.



Our calculations agree well with observations, especially for stars with near-solar values of effective temperature. In particular, our model explains the observed transition between faculae- and spot-dominated variability and highlights the importance of the inclination angle for solar-stellar comparisons. We construe this as a confirmation that our model properly reproduces the relation between spot and facular coverages not only for the Sun but also for more active stars.

\section{Discussion and Conclusions}
All in all, the ability of our model to reproduce both solar and stellar data allows discussing a 
physical mechanism in charge of the decrease of the faculae-to-spot ratio with magnetic activity. 
The instantaneous ratio (i.e. observed at a given time) between facular and spot area coverages depends on (a) the ratio between those parts of a newly emerged BMR attributed to faculae and spots (``driver ratio''), as well as on (b) the decay rates of faculae and spots. The size distribution of the BMRs in our model is fixed (with a slight dependence on the cycle phase \citep{Isik2018}). With increasing activity level (i.e. higher BMR emergence frequency) this distribution is still preserved (see Methods), which means that the driver ratio in our model is assumed to be independent on the activity level.  The spots decay linearly in time, while the decay rate remains the same for all spots, such that the Gnevyshev–Waldmeier relation \citep{Waldmeier1955} between sunspot sizes and lifetimes is satisfied. The area coverage by faculae is generally larger than those of spots (see Figs. \ref{fig:fig_1} and \ref{fig:fig_2}). This in turn means that cancellation of fluxes of opposite polarities,  both directly upon emergence and during the subsequent evolution of surface flux are more likely. As the activity level increases, opposite-polarity facular flux elements lie close to each other more frequently. This leads to more efficient flux cancellation, resulting in shortening of the facular lifetime. Consequently, a more efficient cancellation of the magnetic flux at higher activity levels leads to the drop of the ratio between facular and spot area coverages (see Appendix D for a simple illustrative experiment). We note that our model also takes into account the cancellation of spots with different polarities. However, since the spot areas are much smaller than those of faculae and their lifetimes are much shorter, such cancellations barely occur.

In our numerical experiments, we assumed that active stars simply display higher amounts of flux emergence than the present Sun. At the same time, more active stars are often younger and rotate faster \citep[although there is a significant scatter in the activity-rotation period dependency, e.g. see][]{Barnes2007}. 
Here we neglect this trend and the associated modification of the surface distribution of the magnetic region emergence \citep{Isik2018}.
We, thus, focused purely on the increasing probability of superposition of emerging  and evolving surface flux elements to isolate this effect. We note that the good agreement between our calculations and solar/stellar data by no means excludes the effect of rotation on magnetic emergences, but merely shows that the cancellation mechanism suggested in the present work plays a decisive role in determining the ratio between facular and spot coverages.

Our study shows that while facular contribution to stellar signals decreases with stellar activity, faculae play a significant role in magnetically-driven signals of low activity stars. This is of particular importance to planet-hosting low activity stars since radial velocity monitoring and transmission spectroscopy observations aimed at characterising planets and their atmospheres, respectively, choose such stars as their main targets. \\ \\ \\

N.-E.N., A.I.S. and K. S. received funding from the European Research Council under the European Union’s Horizon 2020 research and innovation program (grant agreement No. 715947). L.G. acknowledges support from ERC Synergy grant WHOLESUN 810218.

\section*{Appendix A}

The SFTM describes the passive transport of the radial
 component of the magnetic field B, considering the effects of differential rotation $\Omega(\lambda)$ (with $\lambda$ being the latitude), meridional flow $\nu(\lambda)$ at the solar surface, and a horizontal surface diffusion thanks to a non-zero horizontal diffusivity $\eta_H$ \citep[50 km$^{2}$s$^{-1}$ taken from][]{Cameron2010}. The emerged active regions gradually disperse due to the radial diffusion $\eta_r$ \citep[25 km$^{2}$s$^{-1}$ as set by][]{Jiang2011_2}, with the flux finally decaying after cancellation between opposite polarities, where they overlap. The governing equation is

\begin{equation}
\begin{split}
  \frac{\partial B}{\partial t} &= - \Omega(\lambda)\frac{\partial B}{\partial \phi} - \frac{1}{R_{\odot}\cos\lambda}\frac{\partial}{\partial \lambda}(\nu(\lambda)B \cos(\lambda))\\&+\eta_H\left ( \frac{1}{R_{\odot}^{2}\cos\lambda}\frac{\partial}{\partial \lambda}\left ( \cos(\lambda)\frac{\partial B}{\partial \lambda} \right ) + \frac{1}{R_{\odot}^{2}\cos^{2}\lambda} \frac{\partial ^{2}B}{\partial\phi^{2}}\right ) \\ &+D(\eta_r) + S(\lambda,\phi,t),
\end{split}
\label{flux}
\end{equation}
\noindent where $R_{\odot}$ is the solar radius, $\phi$ is the longitude of the 
active region, and $D$ is a linear operator that describes the decay due to radial
diffusion with the radial surface diffusivity $\eta_r$ \citep{Baumann2006}.
The time-averaged (synodic) differential rotation profile was taken from \cite{Snodgrass1983} and the time-averaged meridional flow is expressed following \cite{vanBallegooijen1998}. The magnetic flux in the model is injected as bipolar magnetic regions (BMRs, with one BMR consisting of two neighbouring flux concentrations with opposite polarities). The emergence of the BMRs is introduced via the source term that determines the position and time of the emerging flux, as well as the flux density of each polarity patch, namely
\begin{equation}
B^{\pm}(\lambda,\phi) = B_{\mathrm{max}}\left(\frac{0.4 \Delta \beta}{\delta}\right)^{2} e^{(-2[1-\cos(\beta_{\pm}(\lambda,\phi))]/\delta^{2})} ,
\label{source2}
\end{equation}
\noindent where $\beta_{\pm}$($\lambda$,$\phi$) is the heliocentric angle between a given point with latitude and longitude ($\lambda$, $\phi$) and the centres of the polarity patches, $\Delta \beta$ is the angular separation between the centres of the two polarities and $\delta$ is the size of the individual polarity patch, taken to be 4$^\circ$. $B_{\mathrm{max}}$ is a scaling factor set to 374 G \citep{Cameron2010,Jiang2011_2}. The source term is compiled following the procedure described in \cite{Cameron2010} with the statistics of solar BMR emergences from \cite{Jiang2011_2}. All in all, the source term represents, statistically, the observed distribution of tilt-angles, sizes, emergence frequency, and emergence latitudes of solar cycle 22, with the longitudes being randomised.

When converting the magnetic flux of the SFTM output from \cite{Isik2018}, three free parameters from \citetalias{Nina1} are used: the linear decay rate of the spots $R_d$, the linear growth rate of the spots $R_g$ and the faculae saturation threshold $B_{sat}$ as defined in \cite{Krivova2003}). As the input record in the present approach is different, we adjusted the free parameters to reproduce the area coverages of \citetalias{Nina1}. This resulted in values of $R_g=$ 600 MSH/day , $R_d=$ 47 MSH/day and $B_{sat}=470$\,G. 

\section*{Appendix B}

We define the time-dependent emergence rate of BMRs on a star $S_{\star}$ as 
$S_{\star}(t) = S_{\odot}(t)\cdot \tilde{s}$, where  $S_{\odot}(t)$ is the-time dependent solar BMR emergence rate of solar cycle 22 and $\tilde{s}$ is a scaling factor, following the notation in ref. \cite{Isik2018}. This means that a star with $\tilde{s}=2$ exhibits two times more BMR emergences compared to the solar case, $\tilde{s}=1$.
As noted by  \cite{Jiang2011_1} and taken into account by \cite{Isik2018}, there is a weak linear dependence of the mean latitude of BMR emergence on the cycle amplitude (e.g. the cycle-integrated number of emergences). We took this relationship into account when modelling a more active solar cycle. At the same time the size distribution of the BMRs of $\tilde{s}=1$ is preserved also for stars with higher BMR emergence frequency. We neglect effects of the rotation on the latitude of emergence in this work. Similar to considering different inclination angles, shifting the activity belt of active-region emergence towards higher latitudes will result in different visible disc areas due to fore-shortening and will affect the intensity contrasts of the magnetic features \citep[see,][]{Shapiro2014, Isik2018}. 


\section*{Appendix C}

Monitoring of stellar Ca II H\&K  emission revealed that the amplitude of the stellar activity cycle in the S-index is proportional to the cycle-averaged value of the S-index \citep{Hall2009}. We use this property to 
generate an S-index time series of active stars (i.e. stronger cycles) by linearly scaling the annual averages of the solar S-index for cycle 22. For each  S-index, we obtained a time series of the annual S-index \citep{Shapiro2014, Anna2020}.
Then we used the polynomial fit from Fig. \ref{fig:poly-fit}  to generate a corresponding  photometric time series and applied linear regression for the variability in Strömgren (b+y)/2 as a function of the annually averaged S-index, to calculate the slope parameter $\Delta[(b+y)/2]/\Delta  S$. We note that while the scaling of the solar S-index using the activity cycle employed here is only an approximation, we do not expect it to have a major effect on our calculations of the regression coefficient.

\section*{Appendix D}
Here we present a simple experiment, which demonstrates the effect of flux cancellation. We prescribe all emerging active regions to have a size of $\Delta \beta=10\degree$, representing the largest regions in our input record and perform the calculations for $\tilde{s}=$1, 16, and 64, with all BMRs emerging at the same moment of time $t=0$. 

\begin{figure}[h!]
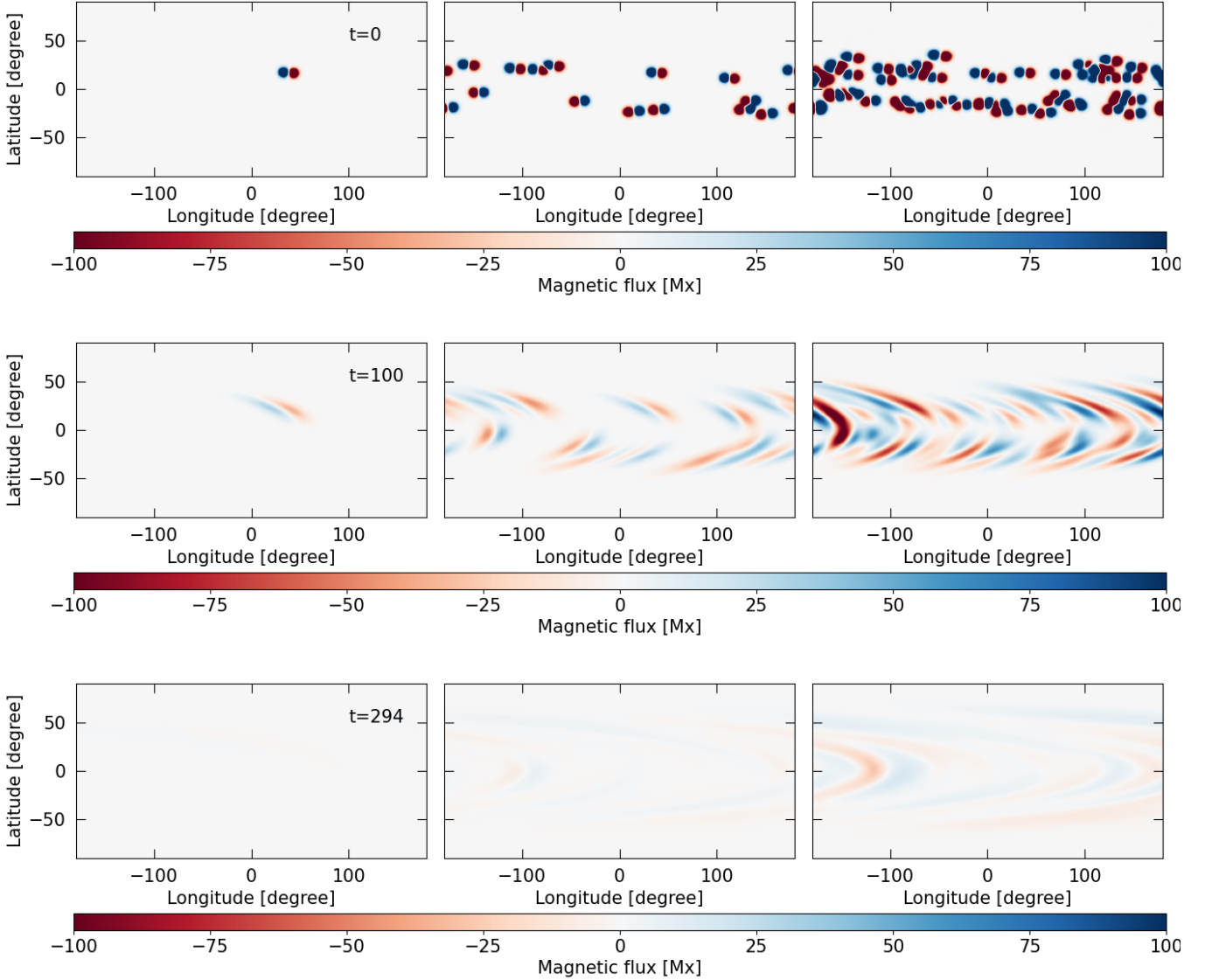

\centering
\includegraphics[width=\textwidth]{Frame005_magnetograms.pdf}\quad
\includegraphics[width=\textwidth]{Frame105_magnetograms.pdf}\quad
\includegraphics[width=\textwidth]{Frame299_magnetograms.pdf}\quad
\caption{Distribution of the magnetic flux associated with faculae in the simple simulation set-up. Shown are simulations for the $\tilde{s}=$ 1 (left columns) $\tilde{s}=$ 16 (middle columns) and  $\tilde{s}=$ 64 (right columns), rows correspond to different timestamps in the available animation.  The video shows the decay of flux from the time of emergence until 295 days after.  The duration of the animation is 17s. The movie can be found here: \url{https://www.dropbox.com/s/ipx7gxqof0vrrol/Supplementary_Movie.mp4?dl=0}.}
\label{fig:sim_mags}
\end{figure}

In Fig.~\ref{fig:sim_mags} we  show the snapshots from the Supplementary Animation illustrating the evolution of magnetic flux for all three cases  considered here. One can see that the cancellation of magnetic flux from different active regions happens both during the emergence (see overlapping active regions in top middle end left) as well as during the subsequent evolution.

\begin{figure}[h!]
    \centering
    \includegraphics[width=\columnwidth]{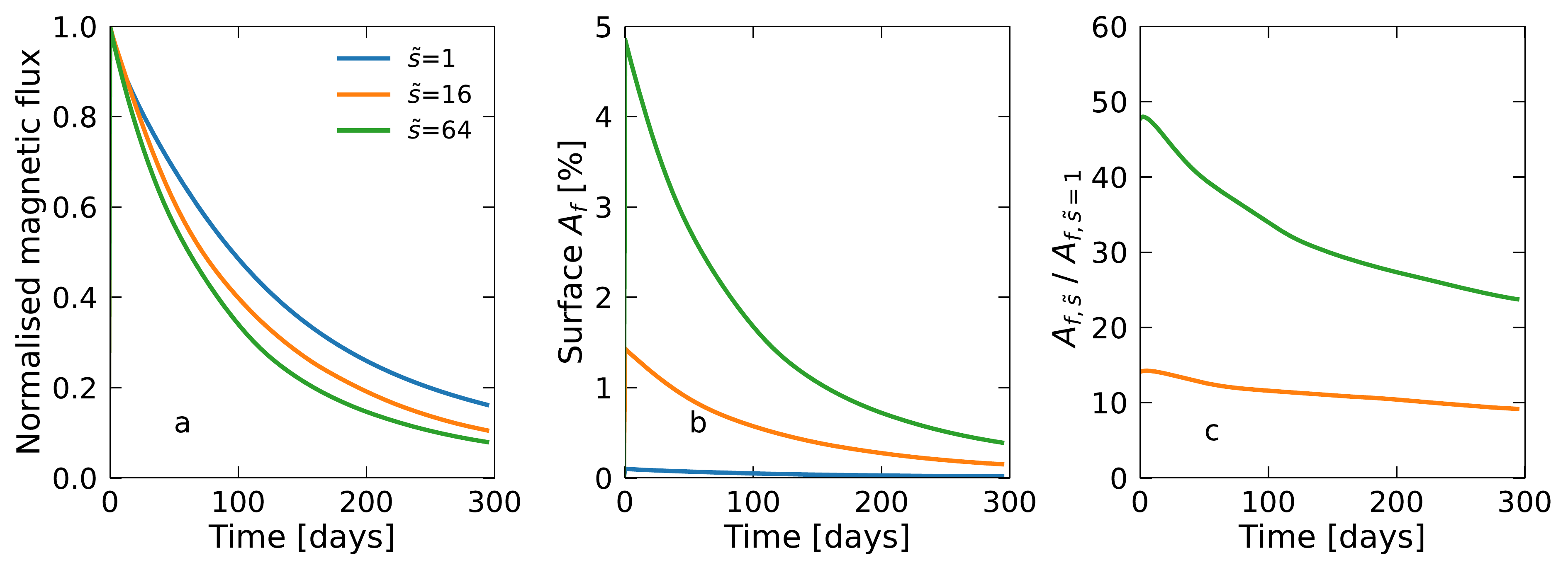}
    \caption{Demonstration of the effect of flux cancellation on the magnetic field and  facular area coverages. Shown are calculations with 1 ($\tilde{s}=$ 1, blue), 16 ($\tilde{s}=$ 16, orange), and 64 ($\tilde{s}=$ 64, orange) active regions. 
    Panel a shows the evolution of the surface magnetic flux normalised to corresponding value at the moment of emergence ($t=0$). Panel b shows the surface area coverage by faculae in percent. Panel c shows the ratios of facular coverages for $\tilde{s}=$ 16 and $\tilde{s}=$ 64 cases to that for $\tilde{s}=$ 1 case (orange and green curves, respectively). Calculations are based on the magnetograms shown in Fig. \ref{fig:sim_mags}.}
    \label{fig:sim_stats}
\end{figure}

Fig.~\ref{fig:sim_stats} shows the time evolution of the magnetic flux (panel a),  the facular area coverages (panel b), and the ratio of facular area coverages for $\tilde{s}=$16 and 64 cases to that for $\tilde{s}=$1 case (panel c).
One can see that both the magnetic flux and facular area coverages  decrease non-linearly with time. Furthermore, the flux drops faster for higher values of $\tilde{s}$ (see Fig.~\ref{fig:sim_stats}a). This implies that the flux cancellation  gets stronger with the increase of the emergence rate and magnetic activity.

We note that the flux cancellation happens not only during the evolution of active regions but also directly at the moment of emergence. Indeed, $A_{f,\tilde{s}=16}$/$A_{f,\tilde{s}=1}$ at  $t=0$ is about 14, while the corresponding $A_{f,\tilde{s}=64}$/$A_{f,\tilde{s}=1}$ is about 50 (see the intersection of orange and green curves with the ordinate $t=0$ in Fig.~\ref{fig:sim_stats}c). Both these values are below the scaling factor for the emergence rate (16 and 64, respectively) so that the surface flux increases slower than the emergence rate, implying  a cancellation of the magnetic flux at the moment of emergence.

\bibliography{bib}
\bibliographystyle{aasjournal}

\end{document}